\documentclass[12pt]{article}
\usepackage{graphicx}

\def\hybrid{\topmargin 0pt      \oddsidemargin 0pt
        \headheight 0pt \headsep 0pt
       \voffset-1cm
        \textwidth 6.25in       
       \textheight 9.5in       
        \marginparwidth 0.0in
        \parskip 5pt plus 1pt   \jot = 1.5ex}
\catcode`\@=11
\def\marginnote#1{}

\newcount\hour
\newcount\minute
\newtoks\amorpm
\hour=\time\divide\hour by60
\minute=\time{\multiply\hour by60 \global\advance\minute by-\hour}
\edef\standardtime{{\ifnum\hour<12 \global\amorpm={am}%
        \else\global\amorpm={pm}\advance\hour by-12 \fi
        \ifnum\hour=0 \hour=12 \fi
        \number\hour:\ifnum\minute<10 0\fi\number\minute\the\amorpm}}
\edef\militarytime{\number\hour:\ifnum\minute<10 0\fi\number\minute}

\def\draftlabel#1{{\@bsphack\if@filesw {\let\thepage\relax
   \xdef\@gtempa{\write\@auxout{\string
      \newlabel{#1}{{\@currentlabel}{\thepage}}}}}\@gtempa
   \if@nobreak \ifvmode\nobreak\fi\fi\fi\@esphack}
        \gdef\@eqnlabel{#1}}
\def\@eqnlabel{}
\def\@vacuum{}
\def\draftmarginnote#1{\marginpar{\raggedright\scriptsize\tt#1}}

\def\draftlabel#1{{\@bsphack\if@filesw {\let\thepage\relax
   \xdef\@gtempa{\write\@auxout{\string
      \newlabel{#1}{{\@currentlabel}{\thepage}}}}}\@gtempa
   \if@nobreak \ifvmode\nobreak\fi\fi\fi\@esphack}
        \gdef\@eqnlabel{#1}}
\def\@eqnlabel{}
\def\@vacuum{}
\def\draftmarginnote#1{\marginpar{\raggedright\scriptsize\tt#1}}

\def\draft{\oddsidemargin -.5truein
        \def\@oddfoot{\sl preliminary draft \hfil
        \rm\thepage\hfil\sl\today\quad\militarytime}
        \let\@evenfoot\@oddfoot \overfullrule 3pt
        \let\label=\draftlabel
        \let\marginnote=\draftmarginnote
   \def\@eqnnum{(\theequation)\rlap{\kern\marginparsep\tt\@eqnlabel}%
\global\let\@eqnlabel\@vacuum}  }

\def\numberbysection{\@addtoreset{equation}{section}
        \def\theequation{\thesection.\arabic{equation}}}

\def\underline#1{\relax\ifmmode\@@underline#1\else
        $\@@underline{\hbox{#1}}$\relax\fi}

\def\titlepage{\@restonecolfalse\if@twocolumn\@restonecoltrue\onecolumn
     \else \newpage \fi \thispagestyle{empty}\c@page\z@
        \def\thefootnote{\fnsymbol{footnote}} }

\def\endtitlepage{\if@restonecol\twocolumn \else  \fi
        \def\thefootnote{\arabic{footnote}}
        \setcounter{footnote}{0}}  
\relax

\hybrid

\newfont{\Bbb}{msbm10 scaled 1\@ptsize00}
\newfont{\Bbbb}{msbm7 scaled 1\@ptsize00}

\newcommand{\DDD}{\raise-1pt\hbox{$\mbox{\Bbbb D}$}}

\newcommand{\PP}{\mbox{\Bbb P}}        

\newcommand{\UUU}{\raise-1pt\hbox{$\mbox{\Bbbb U}$}}

\newcommand{\ZZ}{\mbox{\Bbb Z}}
\newcommand{\z}{\raise-1pt\hbox{$\mbox{\Bbbb Z}$}}

\def\beq{\begin{equation}}
\def\eeq{\end{equation}}
\def\p{\partial}

\begin{document}

\begin{titlepage}

\title{Dispersionless DKP hierarchy and elliptic L\"owner equation}

\author{V.~Akhmedova\thanks{National Research University Higher School of 
Economics, International Laboratory of Representation 
Theory and Mathematical Physics, 
20 Myasnitskaya Ulitsa, Moscow 101000, Russia,
e-mail: valeria-58@yandex.ru}
\and A.~Zabrodin
\thanks{Institute of Biochemical Physics,
4 Kosygina st., Moscow 119334, Russia; ITEP, 25
B.Cheremushkinskaya, Moscow 117218, Russia and
National Research University Higher School of Economics,
International Laboratory of Representation 
Theory and Mathematical Physics,
20 Myasnitskaya Ulitsa,
Moscow 101000, Russia, e-mail: zabrodin@itep.ru}}

\date{April 2014}
\maketitle

\vspace{-7cm} \centerline{ \hfill ITEP-TH-08/14}\vspace{7cm}

\begin{abstract}

We show that the dispersionless DKP
hierarchy (the dispersionless limit of the Pfaff lattice)
admits a suggestive reformulation through elliptic functions.
We also consider one-variable reductions of the dispersionless DKP
hierarchy and show that they are described by an elliptic version
of the L\"owner equation. With a particular choice of the
driving function, the latter appears to be closely related to 
the Painlev\'e VI equation with special choice of parameters.

\end{abstract}

\end{titlepage}

\vspace{5mm}

%

\tableofcontents


\section{Introduction}

The DKP hierarchy is one of the integrable hierarchies with 
$D_{\infty}$ symmetries introduced by 
M.Jimbo and T.Miwa in 1983 \cite{JimboMiwa}.
It was subsequently rediscovered and came to be
also known as the coupled KP hierarchy \cite{HO} and
the Pfaff lattice \cite{AHM,ASM}, see also
\cite{Kakei,IWS,Willox}. The latter name is motivated by the fact 
that some solutions to the hierarchy are expressed through Pfaffians. 
The solutions and the algebraic structure were studied in 
\cite{Kodama,Kodama1,AKM}, the relation to matrix integrals 
was elaborated in \cite{AHM,ASM,Kakei,Orlov}.
Bearing certain similarities with the KP and Toda chain hierarchies,
the DKP one is essentially different and less well understood.

The dispersionless version of the DKP hierarchy (the dDKP
hierarchy) was suggested in 
\cite{Takasaki07,Takasaki09}.
It is an infinite system
of differential equations
\beq\label{D1}
e^{D(z)D(\zeta )F}\left (1-\frac{1}{z^2\zeta^2}\,
e^{2\p_{t_0}(2\p_{t_0} + D(z) + D(\zeta ))F}\right )=
1-\frac{\p_{t_1}D(z)F -\p_{t_1}D(\zeta )F}{z-\zeta}
\eeq
\beq\label{D2}
e^{-D(z)D(\zeta )F}
\, \frac{z^2 e^{-2\p_{t_0}D(z)F}-\zeta^2 e^{-2\p_{t_0}D(\zeta )F}}{z-\zeta}
=z+\zeta -\p_{t_1}\! \Bigl (2\p_{t_0} +D(z)+D(\zeta )\Bigr )F
\eeq
for the function $F=F({\bf t})$ of the infinite number of 
(real) ``times''
${\bf t}=\{t_0, t_1, t_2, \ldots \}$,
where 
\beq\label{D3}
D(z)=\sum_{k\geq 1}\frac{z^{-k}}{k}\, \p_{t_k}\,.
\eeq
The differential equations are obtained by expanding 
equations (\ref{D1}), (\ref{D2}) in powers of $z$, $\zeta$.
For example, the first two equations of the hierarchy are
$$
\left \{ 
\begin{array}{l}
6F_{11}^2 +3F_{22}-4F_{13}=12e^{4F_{00}}
\\ \\
2F_{03}+4F_{01}^3 +6 F_{01}F_{11}-6F_{01}F_{02}=3F_{12}.
\end{array}
\right.
$$
Here and below we use the short-hand notation
$F_{mn}\equiv \p_{t_m}\p_{t_n}F$.
Note that the first equation with $0$ in the right hand side, 
i.e., $6F_{11}^2 +3F_{22}-4F_{13}=0$, is the dispersionless KP 
(Khokhlov-Zabolotskaya) equation written in the Hirota form.

In this paper we study the dDKP hierarchy.
The aim of the paper is two-fold. 

First, we show that somewhat 
unsightly looking equations (\ref{D1}), (\ref{D2}),
when rewritten in an elliptic parametrization in terms of
Jacobi's theta-functions $\theta_a(u|\tau )$, assume a nice 
and suggestive form which looks like a natural elliptic extension
of the dispersionless KP (dKP) hierarchy:
\beq\label{DD4}
(z^{-1}-\zeta^{-1})e^{(\p_{t_0}+D(z))(\p_{t_0}+D(\zeta ))F}
=\frac{\theta_1(u(z)\! -\! 
u(\zeta )|\tau )}{\theta_4(u(z)\! -\! u(\zeta )|\tau )}\,.
\eeq
Here the function $u(z)$ is defined by
\beq\label{DD4a}
e^{\p_{t_0}(\p_{t_0}+D(z))F}=z \, 
\frac{\theta_1(u(z)|\tau )}{\theta_4(u(z)|\tau )}\,.
\eeq
The modular parameter $\tau$  
is a dynamical variable: 
$\tau = \tau ({\bf t})$. This feature suggests some 
similarities with the genus 1 Whitham equations \cite{Krichever}
and the integrable structures behind boundary value 
problems in plane
doubly-connected domains \cite{KMZ05}.

Second, we investigate one-variable reductions of the dDKP 
hierarchy assuming that all dynamical variables depend 
on the times ${\bf t}$
through a single variable which in a generic case can be 
identified with the modular parameter $\tau$.
We show that such reductions are classified by solutions of 
a differential equation which is an elliptic analogue of the 
famous L\"owner equation (see, e.g., 
\cite[Chapter 6]{Pommerenke}). In complex analysis, this 
``elliptic L\"owner equation'' is also known
as the Goluzin-Komatu equation \cite{Goluzin,Komatu}, 
see also \cite{Alexandrov,CDMG1,CDMG2,review}:
\beq\label{DD5}
\begin{array}{l}
4\pi i \, \p_{\tau}u(z, \tau )=-\, E^{(1)}\Bigl ( u(z, \tau ) \! +\! 
\xi (\tau )\, \bigl | \, \frac{\tau}{2}\Bigr )+
\, E^{(1)}\Bigl ( 
\xi (\tau )\, \bigl | \, \frac{\tau}{2}\Bigr ),
\end{array}
\eeq
where $E^{(1)}(u, \tau ):=\p_u \log \theta_1(u|\tau )$ and
$\xi (\tau )$ is an arbitrary (continuous) function of $\tau$ 
(the ``driving function'').
This equation is the basic element of the 
theory of parametric conformal maps from
doubly connected slit domains to annuli. During the last decade,
the interest to this topic was renewed in connection with 
the Schramm-L\"owner evolution (SLE); for the SLE in an annulus see 
\cite{BF,Zhan}. A similar relation between the chordal L\"owner
equation and one-variable 
reductions of the dKP hierarchy was known since the seminal papers
by J.Gibbons and S.Tsarev \cite{GT1,GT2}. 
Further developments are discussed in
\cite{Manas1}-\cite{T13}.

Finally, we point out an unexpected 
connection with the Painlev\'e VI equation. 
Namely, we show that the second 
$\tau$-derivative of the elliptic L\"owner equation 
(\ref{DD5}), with a particular choice of the driving function,
gives the Painlev\'e VI equation with special values of the 
parameters written in the elliptic (``Calogero-like'') form.

\section{The dispersionless DKP hierarchy}

\subsection{Algebraic formulation}

In what follows we will use the differential operator
\beq\label{E5}
\nabla (z)=\p_{t_0}+ D(z)
\eeq
which in the dDKP case is more convenient than $D(z)$. 
Introducing the functions
\beq\label{D4}
p(z)=z-\p_{t_1} \nabla (z)F,
\qquad
w(z)=z^2 e^{-2\p_{t_0}\nabla (z)F},
\eeq
we can rewrite equations (\ref{D1}), (\ref{D2}) in a more compact form
\beq\label{D1a}
e^{D(z)D(\zeta )F}\left (1-\frac{1}{w(z)w(\zeta )}\right )=
\frac{p(z)-p(\zeta )}{z-\zeta}\,,
\eeq
\beq\label{D2a}
e^{-D(z)D(\zeta )F + 2\p_{t_0}^2F}
\, \, \frac{w(z)-w(\zeta )}{z-\zeta}=p(z)+p(\zeta ).
\eeq
Multiplying the two equations, we get the relation
$$
p^2(z)-e^{2F_{00}}\Bigl (w(z)+w^{-1}(z)\Bigr )=
p^2(\zeta )-e^{2F_{00}}\Bigl (w(\zeta )+w^{-1}(\zeta )\Bigr )
$$
from which it follows that $p^2(z)-e^{2F_{00}}\Bigl (w(z)+w^{-1}(z)\Bigr )$
does not depend on $z$. Tending $z$ to infinity, we find that this 
expression is equal to 
$F_{02} -2F_{11} -F_{01}^2$.
Therefore, we conclude that $p(z), w(z)$ satisfy the
algebraic equation \cite{Takasaki09}
\beq\label{D5}
p^2(z)=R^2 \Bigl (w(z)+w^{-1}(z)\Bigr )+V\,,
\eeq
where
\beq\label{D6}
R=e^{F_{00}}, \qquad 
V=F_{02} -2F_{11} -F^2_{01}.
\eeq
This equation defines an elliptic curve, with
$w$, $p$ being algebraic functions on this curve. 
The functions $w$ and $p$ have respectively a double pole 
and a simple pole at infinity.

\subsection{Elliptic formulation}

A natural further step is to uniformize the curve through  
elliptic functions. To this end, we use the standard Jacobi
theta-functions $\theta_a (u)=\theta_a (u|\tau )$ ($a=1,2,3,4$).
Their definition and basic properties are listed in the appendix.

The elliptic parametrization of (\ref{D5}) is as follows:
\beq\label{E1}
w(z)=\frac{\theta_4^2(u(z))}{\theta_1^2(u(z))}\,, \qquad
p(z)=\gamma \, \theta_4^2(0)\, \frac{\theta_2(u(z))\,
\theta_3(u(z))}{\theta_1(u(z))\, \theta_4(u(z))}\,,
\eeq
where $u(z)=u(z, {\bf t})$ is some function 
of $z$ and $\gamma$ is a $z$-independent
factor, and
\beq\label{E2}
R=\gamma\, \theta_2(0)\, \theta_3(0)\,, \qquad
V=-\gamma^2 \Bigl ( \theta_2^4(0)+\theta_3^4(0)\Bigr ).
\eeq
At this stage $\gamma$ is an arbitrary parameter but we will see that
it can not be put equal to a fixed number like 1 because it is a 
dynamical variable, as well as the modular parameter $\tau$:
$\gamma =\gamma ({\bf t})$, $\tau =\tau ({\bf t})$.
In this parametrization, the equation of the curve is equivalent to
the identity
$$
\theta_4^2(0)\, \frac{\theta_2^2(u)\,
\theta_3^2(u)}{\theta_1^2(u)\, \theta_4^2(u)}=
\theta_2^2(0)\theta_3^2(0)\! \left (\frac{\theta_4^2(u)}{\theta_1^2(u)}+
\frac{\theta_1^2(u)}{\theta_4^2(u)}\right )-
\Bigl (\theta_2^4(0)+\theta_3^4(0)\Bigr )
$$
which can be proved either by using some standard identities
for theta-functions or by comparing analytical properties 
of the both sides.
It is convenient to normalize $u(z)$ by the condition 
$u(\infty )=0$, then the expansion around $\infty$ is
\beq\label{E3}
u(z, {\bf t})=\frac{c_1({\bf t})}{z}+\frac{c_2({\bf t})}{z^2}+\ldots 
\eeq

It is not difficult to check the identity
$$
\frac{w(z_1)-w(z_2)}{p(z_1)+p(z_2)}=-\frac{1}{\gamma \,
\theta_2(0)\theta_3(0)}\,\,
\frac{\theta_4(u_1)\theta_4(u_2)}{\theta_1(u_1)\theta_1(u_2)}\,\,
\frac{\theta_1(u_1-u_2)}{\theta_4(u_1-u_2)},
$$
where $u_i\equiv u(z_i)$. This identity allows one to represent
equations (\ref{D1a}), (\ref{D2a}) as a single
equation:
\beq\label{E4}
\left (z_1^{-1}-z_2^{-1}\right ) e^{\nabla (z_1) \nabla (z_2)F}
=\frac{\theta_1(u(z_1)\! -\! u(z_2))}{\theta_4(u(z_1)\! -\! u(z_2))}\,.
\eeq
Note that the limit $z_2\to \infty$ in (\ref{E4}) 
gives the definition of the 
function $u(z)$:
\beq\label{E6}
e^{\p_{t_0}\nabla (z)F}=z\,
\frac{\theta_1(u(z))}{\theta_4(u(z))}
\eeq
(equivalent to the first formula in (\ref{E1})).
In addition, we see from (\ref{E2}) that
\beq\label{E7}
- \frac{V}{R^2}=\, e^{-2F_{00}}\! \left (
2F_{11} \! +\! F^2_{01} \! -\! F_{02}\right )\, 
=\,
\frac{\theta_2^2(0|\tau )}{\theta_3^2(0|\tau )}+
\frac{\theta_3^2(0|\tau )}{\theta_2^2(0|\tau )}\,.
\eeq
The $z\to \infty$ limit of equation (\ref{E6}) yields:
\beq\label{E8}
e^{F_{00}}=R=\pi c_1 \, \theta_2(0)\theta_3(0),
\eeq
hence
\beq\label{E9}
c_1({\bf t})=\frac{\gamma ({\bf t})}{\pi}\,.
\eeq

Yet another useful form of equation (\ref{E4}) can be obtained
by passing to logarithms and applying $\p_{t_0}$ to both sides. 
It is convenient to introduce the function
\beq\label{E10}
S(u| \, \tau ):=\log \frac{\theta_1(u |\tau )}{\theta_4(u |\tau )}\,,
\eeq
which has the following quasiperiodicity properties:
\beq\label{E10a}
S(u+1|\tau )=S(u|\tau )+i\pi \,, \qquad
S(u+\tau |\tau )=S(u|\tau ).
\eeq
In terms of this function, the equation reads
\beq\label{E11}
\nabla (z_1) S\Bigl (u(z_2)|  \tau\Bigr ) =
\p_{t_0}S\Bigl (u(z_1) \! -\! u(z_2)| \tau \Bigr ).
\eeq
In particular, this equation means that the left hand side 
is symmetric with respect to the permutation $z_1 \leftrightarrow z_2$:
$\nabla (z_1) S\Bigl (u(z_2)| \tau\Bigr ) =
\nabla (z_2) S\Bigl (u(z_1)| \tau\Bigr )$. This symmetry 
is a manifestation of integrability.
In the limit $z_2\to \infty$ equation (\ref{E11}) gives:
\beq\label{E12}
\nabla (z)\log R= \p_{t_0} S\Bigl (u(z)\bigr | \tau\Bigr ).
\eeq
In order to connect this with the algebraic formulation, we note that
\beq\label{E13}
S(u(z)|\tau )= -\frac{1}{2}\, \log w(z), \qquad
c_1 S'(u(z)|\tau )= p(z),
\eeq
where $S'(u|\tau )\equiv \p_u S(u|\tau )$. The first formula directly
follows from the definitions. To derive the second one, 
we use equation (\ref{Sprime}) from the appendix.

\section{One-variable reductions}

One may look for solutions of the hierarchy such that 
$u(z, {\bf t})$ and $\tau ({\bf t})$ depend on the times 
through a single variable $\lambda = \lambda ({\bf t})$:
$u(z, {\bf t})=u(z, \lambda ({\bf t}))$, 
$\tau ({\bf t})=\tau (\lambda ({\bf t}))$. 
Such solutions are called one-variable reductions of the hierarchy. 
The function
of two variables, $u(z, \lambda )$, can not be arbitrary.
Our next goal is to characterize the class of functions 
$u(z, \lambda )$, $\tau (\lambda )$ that are consistent with
the structure of the hierarchy and can be used for one-variable 
reductions.

\subsection{The consistency condition for one-variable reductions}

Applying the chain rule of differentiation to 
$
S\Bigl (u(z, {\bf t})| \tau ({\bf t})\Bigr )=
S \Bigl ( u(z, \lambda ({\bf t}))\, |\, \tau (\lambda ({\bf t}))\Bigr )
$,
we get from equation (\ref{E11}):
$$
\nabla (z_1)S(u(z_2))=[\nabla (z_1)\lambda ]\,
\Bigl (\p_{\lambda }u(z_2)S'(u(z_2))+\p_{\lambda }\tau \dot S(u(z_2))\Bigr ).
$$
Hereafter, we write simply $u(z):=u(z, \lambda )$,
$S(u):=S(u|\tau )$ and denote $S'(u)=\p_u S(u|\tau )$,
$\dot S(u)=\p_{\tau} S(u|\tau )$.
Next, we have, using (\ref{E12}):
$$
\nabla (z_1)\lambda = \frac{d\lambda}{d\log R}\, \nabla (z_1)\log R
=\frac{d\lambda}{d\log R}\, \p_{t_0}S(u(z_1))
$$
$$
=\,\, \frac{d\lambda}{d\log R}\, \p_{t_0}\lambda 
\Bigl ( \p_{\lambda }u(z_1)S'(u(z_1))+\p_{\lambda }\tau \dot S(u(z_1))\Bigr ).
$$
In a similar way, we get:
$$
\p_{t_0}S\Bigl (u(z_1)\! -\! u(z_2)\Bigr )=\p_{t_0}\lambda 
\Bigl [
\Bigl (\p_{\lambda}u(z_1)\! -\! \p_{\lambda}u(z_2)\Bigr )
S'\Bigl (u(z_1)\! -\! u(z_2)\Bigr ) +\p_{\lambda }\tau \dot S
\Bigl (u(z_1)\! -\! u(z_2)\Bigr ) \Bigr ].
$$
The formulas simplify a bit if we choose $\lambda =\tau$.
Assuming that $\p_{t_0}\tau$ is not identically zero, we arrive at
the following relation:
\beq\label{one1}
\begin{array}{c}
\Bigl [\p_{\tau}u(z_1)S'(u(z_1))+\dot S(u(z_1))\Bigr ]\,
\Bigl [\p_{\tau}u(z_2)S'(u(z_2))+\dot S(u(z_2))\Bigr ]
\\ \\
=\,\, \displaystyle{
\frac{d\log R}{d\tau}\Bigl [
\Bigl (\p_{\tau}u(z_1)\! -\! \p_{\tau}u(z_2)\Bigr )S' 
\Bigl (u(z_1)\! -\! u(z_2)\Bigr )
+\dot S  \Bigl (u(z_1)\! -\! u(z_2)\Bigr )\Bigr ].}
\end{array}
\eeq
Note that this relation can be written in the compact form
\beq\label{one2}
\frac{d  S(u(z_1))}{d\tau }\,\frac{d  S(u(z_2))}{d\tau }\, =
\, \frac{d\log R}{d\tau} \, 
\frac{d  S \Bigl (u(z_1)\! - \! u(z_2)\Bigr )}{d\tau }\,,
\eeq
where $d/d\tau$ is the total $\tau$-derivative. 

To proceed, we need to know $\dot S(u)$. 
It is given by the formula
\beq\label{one3}
2\pi i \, \dot S(u)=S'(u)\, E^{(2)}(u) +\frac{\pi^2}{2}\,
\theta_4^4(0)
\eeq
which is proved in the appendix. Here and below we use the 
notation\footnote{Note that the standard notation for the 
Eisenstein function $E^{(1)}$ is $E_1$.}
$$
E^{(a)}(u)=E^{(a)}(u |\tau ) = \p_u \log \theta_a(u |\tau ).
$$
The properties of these Eisenstein-like functions 
that we need for calculations are listed
in the appendix (see (\ref{pr1}), (\ref{pr2})).
Using (\ref{one3}), we rewrite (\ref{one1}) in the form
\beq\label{one4}
\begin{array}{c}
\displaystyle{
S'(u_1)\left [4\pi i \, \p_{\tau}u_1 \! + \! 2 E^{(2)}(u_1)\!  +\!
\frac{\pi^2 \theta_4^4(0)}{S'(u_1)}\right ] 
S'(u_2)
\left [4\pi i \, \p_{\tau}u_2 \! + \! 2E^{(2)}(u_2) \! +\! 
\frac{\pi^2 \theta_4^4(0)}{S'(u_2)}\right ]}
\\ \\
=\, 4\pi i \, \displaystyle{
\frac{d\log R}{d\tau}\, S'(u_1\! -\! u_2)
\left [4\pi i (\p_{\tau}u_1 \! -\! \p_{\tau}u_2)
\! +\! 2 E^{(2)}(u_1\! -\! u_2) \! +\! 
\frac{\pi^2 \theta_4^4(0)}{S'(u_1\! -\! u_2)}\right ],}
\end{array}
\eeq
where $u_j\equiv u(z_j)$ for brevity. Now, one can see that 
the substitutions
\beq\label{one5}
\left \{
\begin{array}{l}
4\pi i \, \p_{\tau}u=-E^{(1)}(u+\xi )-E^{(4)}(u+\xi )+
E^{(1)}(\xi )+E^{(4)}(\xi ),
\\ \\
4\pi i \, \p_{\tau}\log R =(S'(\xi ))^2,
\end{array} \right.
\eeq
where $\xi$ is an arbitrary parameter, convert equation 
(\ref{one4}) into identity. Some details are given in the appendix.
This means that the function $u(z, \tau )$ is compatible 
with the infinite hierarchy if it satisfies the differential equation
\beq\label{one6}
4\pi i \, \p_{\tau}u(z)=-E^{(1)}(u(z)+\xi(\tau ) |\tau )-
E^{(4)}(u(z)+\xi (\tau ) |\tau  )+
E^{(1)}(\xi (\tau ) |\tau  )+E^{(4)}(\xi (\tau )|\tau  ),
\eeq
where $\xi (\tau)$ can be arbitrary function of $\tau$. 
The identity $E^{(1)}(u|\tau )+E^{(4)}(u|\tau )=
E^{(1)}(u|\frac{\tau}{2} )$ allows one to write this equation 
in a more compact form:
\beq\label{one6a}
4\pi i \, \p_{\tau}u(z)=-E^{(1)}\Bigl (u(z)+\xi (\tau )
|\, \frac{\tau}{2}\Bigr )+
E^{(1)}\Bigl (\xi (\tau )|\, \frac{\tau}{2}\Bigr ).
\eeq
This is the elliptic analogue of the L\"owner equation 
known also as the Goluzin-Komatu equation \cite{Goluzin,Komatu}.
One can also see that the equation 
\beq\label{one7}
4\pi i \, \p_{\tau}\log R =(S'(\xi (\tau )))^2
\eeq
emerges as the limiting case of (\ref{one6}) when $z\to \infty$.
The function $\xi (\tau )$ is
the ``driving function'' that encodes the shape of the slit
in the L\"owner theory. In our setting, it specifies
the reduction.

Two remarks are in order.
\begin{itemize}
\item[a)]
Using the identity proved in the appendix, it is possible to show that
for one-variable reductions the total $\tau$-derivative of $S(u(z))$
is given by
\beq\label{one8}
4\pi i \frac{dS(u(z))}{d\tau}=S'(\xi (\tau ))\, 
S'\Bigl (u(z)+\xi (\tau )\Bigr ).
\eeq
\item[b)] The $u(z)$-independent second term in the right hand side of
equation (\ref{one6a}) can be eliminated by another 
choice of normalization. Indeed,
let us consider the function $\tilde u(z)=u(z)+c_0 (\tau )$, i.e.,
$$
\tilde u(z)=c_0 (\tau )+\frac{c_1(\tau )}{z}+
\frac{c_2(\tau )}{z^{2}}+ \ldots  \quad \mbox{with} \quad
c_0 (\tau )=-\frac{1}{4\pi i}\int_{\tau_0}^{\tau}
E^{(1)}\Bigl (\xi (\tau ')\Bigr |\frac{\tau '}{2}\Bigr )d\tau '
$$
and set $\tilde \xi = \xi -c_0$. Then the elliptic L\"owner equation 
(\ref{one6a}) acquires the form
\beq\label{one6b}
4\pi i \, \p_{\tau}\tilde u(z)=-E^{(1)}\Bigl (\tilde u(z)+\tilde \xi (\tau )
\, \Bigr |\, \frac{\tau}{2}\Bigr ).
\eeq

\end{itemize}

\subsection{The system of reduced equations and their solution}

In order to complete the description of one-variable reductions,
we should derive the equation satisfied by $\tau ({\bf t})$ and 
find its solution. Following the way we have used to derive relation
(\ref{one2}), we write:
$$
\nabla (z) \tau = 
\frac{\p_{\tau}u(z) S'(u(z))+\dot S (u(z))}{d\log R/d\tau}\, \,
\p_{t_0}\tau = \frac{d S(u(z))/d\tau}{d\log R/d\tau}\, \,
\p_{t_0}\tau .
$$
Substituting (\ref{one7}) and (\ref{one8}), we get:
\beq\label{one9}
\nabla (z) \tau = \frac{S'\Bigl (u(z)+\xi (\tau )\Bigr )}{S'(\xi (\tau ))}
\, \, \p_{t_0}\tau .
\eeq
This is a generating equation for a hierarchy of equations 
of the hydrodynamic type. To write them explicitly, we use the expansion
\beq\label{one10}
S'(u(z)+u)=S'(u)+ \sum_{k\geq 1}\frac{z^{-k}}{k}\, B'_k(u)
\eeq
which defines the functions $B_k'(u)=B_k'(u|\tau )$. 
In terms of these functions,
the equations of the reduced hierarchy are as follows:
\beq\label{one11}
\frac{\p \tau}{\p t_k}= \phi_k (\xi (\tau )|\tau )\, 
\frac{\p \tau}{\p t_0}\,, \qquad
\phi_k (\xi (\tau )|\tau ):=
\frac{B'_k(\xi (\tau )|\tau)}{S'(\xi (\tau )|\tau)}\,, \quad
k\geq 1.
\eeq
The common solution to these equations can be written in the 
hodograph form:
\beq\label{one12}
\sum_{k=1}^{\infty} t_k \phi_k(\xi (\tau )|\tau)=\Phi (\tau ),
\eeq
where $\Phi (\tau )$ is an arbitrary function of $\tau$. In
the simplest case, when $\Phi (\tau )=0$, we conclude from (\ref{one12}) 
that 
$\displaystyle{
\sum_{k\geq 1}t_k \, \frac{\p \tau}{\p t_k}\, =0}
$,
i.e., $\tau ({\bf t})$ is a homogeneous function of the times 
of degree $0$.

\subsection{A connection with Painlev\'e VI}

Here we work with the elliptic L\"owner 
equation in the normalization (\ref{one6b}) skipping 
tilde from the notation and changing $\tau \to 2\tau$:
\beq\label{one6c}
2\pi i \, \p_{\tau} u(z)=-E^{(1)}\Bigl (u(z)+\xi 
\, \Bigr |\, \tau \Bigr ).
\eeq
As an example, consider the simplest possible case 
when $\xi$ does not depend on $\tau$: $\xi =\mbox{const}$
(in the dKP case, such a choice of the driving function 
means the reduction to the dispersionless
KdV hierarchy or hierarchies equivalent to it).
Assume that $u(z)=u(z, \tau )$ satisfies this equation.
An easy calculation with the use of (\ref{pr3}) shows that
the function $f(\xi , \tau ):=E^{(1)}\Bigl (u(z, \tau )+\xi |\tau \Bigr )$ 
obeys the
heat equation\footnote{We thank A.Levin who conjectured this fact.}
\beq\label{heat1}
4\pi i \, \p_{\tau} f(\xi , \tau )=\p_{\xi}^2 f(\xi , \tau ).
\eeq
Applying $\p_{\tau}$ to the both sides of equation (\ref{one6c}),
we get, using the heat equation (\ref{heat1}):
\beq\label{ex1}
(2\pi i)^2 \p_{\tau}^2 u=
\frac{1}{2}\, \wp '(u+\xi ),
\eeq
where $\wp (u)=
-\p_uE^{(1)}(u)+\mbox{const}$ is the Weierstrass $\wp$-function with periods 
$1$ and $\tau$. If $\xi =0$ or $\xi =\frac{1}{2}$, this is 
the Painlev\'e VI equation written in the elliptic form 
with a special choice of the parameters \cite{Manin}.

\section{Concluding remarks}

We have demonstrated that the Pfaff lattice (an infinite integrable
hierarchy with the $D_{\infty}$ symmetry)
in the dispersionless limit can be naturally reformulated 
as an ``elliptic deformation'' of the usual dKP hierarchy.
This seems to be a rather surprising and somewhat
mysterious fact. What could be the hidden link
between Pfaffians and elliptic functions?

Once the elliptic reformulation has been done, 
the description of one-variable reductions of the 
dDKP hierarchy obtained in this paper looks rather natural
if one keeps in mind the corresponding 
Gibbons-Tsarev result for the dKP case. 
To wit, the one-variable reductions, i.e., reductions with only one 
independent function, are obtained from solutions to the 
elliptic analogue of the L\"owner equation (the Goluzin-Komatu 
equation), well known in the theory of conformal maps of
doubly-connected slit domains. We hope to 
clarify the geometric meaning of the 
reductions, and of the hierarchy in general, in subsequent 
publications. 

It should be noted that we have found only sufficient conditions
for the consistent one-variable reductions. In order to find 
the necessary conditions and to give a complete description, 
one should find all solutions to  
the functional relation (\ref{one2}), which is the consistency 
condition for the reductions.  

A more complicated problem is to describe 
multi-variable reductions.
Here one can anticipate that an elliptic analogue
of the system of the Gibbons-Tsarev equations should come into play
as consistency conditions. 

An unexpected observation, which seems to be especially interesting, 
is the close connection with the Painlev\'e VI equation.
For a particular (simplest possible?) choice of the driving function,
the elliptic L\"owner equation appears to be the integrated
Painlev\'e VI with special values of the parameters.
Note that in the dKP case, the simplest possible driving function 
(equal to zero) corresponds 
to the most familiar and explicit reduction, the one to the dispersionless 
KdV hierarchy. This is one of the very few cases when the 
chordal L\"owner equation 
can be explicitly solved.
It would be very interesting to find such solvable cases 
for the elliptic version of the L\"owner equation.

\section*{Appendix}
\addcontentsline{toc}{section}{Appendix}
\def\theequation{A\arabic{equation}}
\setcounter{equation}{0}

\subsection*{Theta-functions}

The Jacobi's theta-functions $\theta_a (u)=
\theta_a (u|\tau )$, $a=1,2,3,4$, are defined by the formulas
\beq\label{Bp1}
\begin{array}{l}
\theta _1(u)=-\displaystyle{\sum _{k\in \z}}
\exp \left (
\pi i \tau (k+\frac{1}{2})^2 +2\pi i
(u+\frac{1}{2})(k+\frac{1}{2})\right ),
\\
\theta _2(u)=\displaystyle{\sum _{k\in \z}}
\exp \left (
\pi i \tau (k+\frac{1}{2})^2 +2\pi i
u(k+\frac{1}{2})\right ),
\\
\theta _3(u)=\displaystyle{\sum _{k\in \z}}
\exp \left (
\pi i \tau k^2 +2\pi i u k \right ),
\\
\theta _4(u)=\displaystyle{\sum _{k\in \z}}
\exp \left (
\pi i \tau k^2 +2\pi i
(u+\frac{1}{2})k\right ),
\end{array}
\eeq where $\tau$ is a complex parameter (the modular parameter) 
such that ${\rm Im}\, \tau >0$. The function 
$\theta_1(u)$ is odd, the other three functions are even.
The infinite product representation for the $\theta_1(u)$ reads: 
\beq
\label{infprod} \theta_1(u)=i\,\mbox{exp}\, 
\Bigl ( \frac{i\pi \tau}{4}-i\pi u\Bigr ) \prod_{k=1}^{\infty} 
\Bigl (
1-e^{2\pi i k\tau }\Bigr ) \Bigl ( 1-e^{2\pi i ((k-1)\tau +u)}\Bigr ) 
\Bigl ( 1-e^{2\pi i (k\tau -u)}\Bigr ). 
\eeq 
We also mention the identity
\beq\label{theta1prime}
\theta_1'(0)=\pi \theta_2(0) \theta_3(0) \theta_4(0).
\eeq

In order
to unify some formulas given below, it is convenient to 
understand the index $a$ modulo $4$, i.e., to identify $\theta_{a} (z) \equiv \theta_{a+4} (z)$. 
Set $ \omega _0 =0$, $\omega_1 =\frac{1}{2}$, 
$\omega _2=\frac{1+\tau}{2}$, $ \omega _3
=\frac{\tau}{2}$ then the function $\theta _a(u)$ 
has simple zeros at the points of the lattice $\omega _{a-1}+\ZZ +\ZZ
\tau $.

The theta-functions have the following quasi-periodic properties under shifts
by $1$ and $\tau$: 
\beq\label{Bp1a}
\begin{array}{l}
\theta _a (u+1)=e^{\pi i(1+2\p_{\tau}\omega_{a\! -\! 1})}
\theta _a(u)
\\
\theta _a (u+\tau )=e^{\pi i(a+2\p_{\tau}\omega_{a\! -\! 1})}
e^{-\pi i\tau -2\pi iu}\theta _a(u).
\end{array}
\eeq
Shifts by the half-periods relate
the different theta-functions to each other:
\beq\label{shifts1}
\theta_1(u+ \omega_1)=\, \theta_2(u)\,,
\quad
\theta_3(u+ \omega_1)=\theta_4(u)\,,
\eeq
\beq\label{shifts2}
\theta_1(u+\omega_2)=e^{-\frac{\pi i \tau}{4}
-\pi i u} \theta_3(u)\,,
\quad
\theta_2(u+ \omega_2)=-ie^{-\frac{\pi i \tau}{4}
-\pi i u} \theta_4(u)
\eeq
\beq\label{shifts3}
\theta_1(u+ \omega_3)=ie^{-\frac{\pi i \tau}{4}
-\pi i u} \theta_4(u)\,,
\quad
\theta_2(u+ \omega_3)=e^{-\frac{\pi i \tau}{4}
-\pi i u} \theta_3(u).
\eeq

In the main text we use the special
notation for the Eisenstein-like functions:
$$
E^{(a)}(u)=E^{(a)}(u |\tau ) = \p_u \log \theta_a(u |\tau ).
$$
Using (\ref{Bp1a}), (\ref{shifts1}), 
it is easy to prove the following properties of the 
functions $E^{(a)}(u)$:
\beq\label{pr1}
E^{(a)}(u+1)=E^{(a)}(u)\,, \qquad
E^{(a)}(u+\tau )=E^{(a)}(u)-2\pi i
\eeq
and
\beq\label{pr2}
\begin{array}{l}
E^{(1)}(u+\frac{\tau}{2})=E^{(4)}(u)-\pi i\,,
\\
E^{(4)}(u+\frac{\tau}{2})=E^{(1)}(u)-\pi i\,.
\end{array}
\eeq
For calculations we also need 
$E^{(2)}(0)=0$, $E^{(2)}(\frac{\tau}{2})=-\pi i$.

All formulas for derivatives of elliptic functions with respect 
to the modular parameter follow from the ``heat equation'' satisfied
by the theta-functions:
\beq\label{heat}
4\pi i \, \p_{\tau}\theta_a(u)=\p_u^2 \theta_a(u).
\eeq
In particular, the $\tau$-derivative 
of the Eisenstein function is given by
\beq\label{pr3}
4\pi i \p_{\tau} E^{(1)}(u|\tau )=2E^{(1)}(u|\tau )E^{(1)}{}'(u|\tau )+
E^{(1)}{}''(u|\tau )
\eeq
(see, e.g., \cite{ZZ12}).

\subsection*{Proof of equation (\ref{one3})}

Here we prove formula (\ref{one3}) for the $\tau$-derivative 
of the function $\displaystyle{S(u|\tau )=
\log \frac{\theta_1(u|\tau )}{\theta_4(u|\tau )}}\,$:
\beq\label{one3a}
2\pi i \, \p_{\tau}S(u|\tau )=\p_u S(u|\tau )\, E^{(2)}(u|\tau ) +
\frac{\pi^2}{2}\,
\theta_4^4(0|\tau )\,.
\eeq
A similar formula has been derived in \cite{Takasaki01,ZZ12}
in the context of the Painlev\'e-Calogero correspondence.

We start with the following factorized representation of $S'(u)$:
\beq\label{Sprime}
S'(u)=E^{(1)}(u)-E^{(4)}(u)=
\pi \theta_4^2(0)\, 
\frac{\theta_2(u)\, \theta_3(u)}{\theta_1(u)\, \theta_4(u)}\,,
\eeq 
which can be easily proved, with the help of 
equation (\ref{theta1prime}), by comparing analytical properties 
of the both sides. We will also need the particular case of this
identity obtained by shifting $u\to u+\frac{1}{2}$, taking the 
$u$-derivative and tending $u\to 0$:
\beq\label{Sprime1}
\frac{\theta_{3}''(0)}{\theta_3(0)}-
\frac{\theta_{2}''(0)}{\theta_2(0)}=\pi^2 \theta_4^4(0).
\eeq

From 
(\ref{E10a}) we see that $\dot S(u+1)=\dot S(u)$, 
$\dot S(u+\tau )=\dot S(u)-S'(u)$, so both sides of equation 
(\ref{one3a}) are periodic under the shift $u\to u+1$ and 
gain the additive contribution $-2\pi i S'(u)$ under the shift $u\to u+\tau$
(see (\ref{pr1})). Therefore, the function
$$
g(u):= 4\pi i \, \dot S(u)-2S'(u)E^{(2)}(u)-\pi^2
\theta_4^4(0)
$$
is doubly-periodic with primitive periods $1$, $\tau$.
Using the heat equation (\ref{heat}), we have:
$$
g(u)=\frac{\theta_{1}''(u)}{\theta_1(u)}-\frac{\theta_{4}''(u)}{\theta_4(u)}
-2\pi \theta_4^2 (0)\, \frac{\theta_3(u)\, \theta_{2}'(u)}{\theta_1(u)\,
\theta_4(u)}-\pi^2 \theta_4^4(0).
$$
In order to prove that $g(u)\equiv 0$, it is enough to show that 
it is regular at $u=0$, $u=\frac{\tau}{2}$ (zeros of the denominators)
and $g(u_0)=0$ at some point $u_0$ (it is convenient 
to choose $u_0=\frac{1+\tau}{2}$). The regularity at $u=0$ is obvious
since $\theta_{1}''(u)$ and 
$\theta_{2}'(u)$ have simple zeros at $u=0$ which cancel zeros
in the denominators. The regularity at $u=\frac{\tau}{2}$ is less 
obvious but holds due to identity (\ref{theta1prime}). Finally,
$g(\frac{1+\tau}{2})$ can be found to be zero 
with the help of (\ref{Sprime1}). Clearly, $g(u)\equiv 0$ 
is equivalent to (\ref{one3a}).

\subsection*{Proof of the key identity}

Here we prove the key identity which allows one to derive the
elliptic L\"owner equation from (\ref{one4}).
Set
$$
\varphi (x_1, x_2):=-E^{(1)}(x_1) -E^{(4)}(x_1) +
E^{(1)}(x_2) +E^{(4)}(x_2) +2E^{(2)}(x_1-x_2).
$$
The identity is
\beq\label{key1}
S'(x_1-x_2)\varphi (x_1, x_2) +\pi^2 \theta_4^4(0)=
S'(x_1)S'(x_2).
\eeq
To prove it, we note that $\varphi (x_1, x_2)$ admits 
the following factorized representation:
\beq\label{key2}
\varphi (x_1, x_2)=\pi \theta_2(0)\theta_3(0)\theta_4^2(0)\,
\frac{\theta_1(x_1\! -\! x_2)\, \theta_4(x_1\! -\! x_2)\, 
\theta_2(x_1\! +\! x_2)}{\theta_1(x_1)\theta_4(x_1)
\theta_1(x_2)\theta_4(x_2)\theta_2(x_1\! -\! x_2)}\,.
\eeq
The proof is standard in the theory of
elliptic functions. We should check that: a) the both sides
are doubly periodic as functions of 
$x_1$ with periods $1$ and $\tau$, b) 
the both sides have the same zeros and poles. Therefore, 
they differ by an $x_1$-independent
factor (actually equal to $1$) which can 
be found by tending $x_1\to 0$.
Next, substitute the explicit form of $S'(x)$ (\ref{Sprime})
to the left hand side of (\ref{key1}). We get:
$$
{\rm LHS}\, = \pi^2 \theta_4^4(0)\left (
\frac{\theta_2(0)\theta_3(0)\, 
\theta_2(x_1\! +\! x_2)\theta_3(x_1\! -\! x_2)}{\theta_1(x_1)\theta_4(x_1)
\theta_1(x_2)\theta_4(x_2)}\, +1\right ).
$$
The same argument as above shows that this function is equal to
$S'(x_1)S'(x_2)$.

\subsection*{Derivation of the elliptic L\"owner equation 
from (\ref{one4})}

We should show that the substitution (\ref{one5}),
\beq\label{one5a}
\left \{
\begin{array}{l}
4\pi i \, \p_{\tau}u=-E^{(1)}(u+\xi )-E^{(4)}(u+\xi )+
E^{(1)}(\xi )+E^{(4)}(\xi ),
\\ \\
4\pi i \, \p_{\tau}\log R =(S'(\xi ))^2,
\end{array} \right.
\eeq
converts 
(\ref{one4}) into identity. Indeed, after this substitution 
equation (\ref{one4}) acquires the form
$$
S'(u_1) \left [\varphi (u_1 \! +\! \xi, \xi )+
\frac{\pi^2\theta_4^4(0)}{S'(u_1)}\right ]
S'(u_2) \left [\varphi (u_2\! +\! \xi, \xi )+
\frac{\pi^2\theta_4^4(0)}{S'(u_2)}\right ]
$$
$$
=\, (S'(\xi ))^2S'(u_1\! -\! u_2) \left [\varphi (u_1\! +\! \xi ,\,
u_2\! +\! \xi )+\frac{\pi^2\theta_4^4(0)}{S'(u_1\! -\! u_2)}\right ].
$$
It remains to employ the identity (\ref{key1}) for 
$(x_1, x_2)=(u_1\! +\! \xi , \,\xi )$, 
$(x_1, x_2)=(u_2\! +\! \xi , \,\xi )$ and
$(x_1, x_2)=(u_1\! +\! \xi , \, u_2\! +\! \xi)$.

Finally, we should check that equation (\ref{one7}),
$$
4\pi i \, \p_{\tau} \log R=(S'(\xi (\tau )))^2,
$$
is the limiting case of (\ref{one6}) as $z\to \infty$.
Substituting the series (\ref{E3}) into (\ref{one6}) and 
comparing the leading terms, we get:
$$
4\pi i \, \p_{\tau} \log c_1 =-E^{(1)}{}'(\xi (\tau ))-
E^{(4)}{}'(\xi (\tau )),
$$
where $E^{(a)}{}'(u)=\p_u E^{(a)}(u)$.
Recall that $\log R=\log (\pi c_1)+ \log \Bigl ( \theta_2(0)
\theta_3(0)\Bigr )$ (see (\ref{E2}), (\ref{E9})), so
$$
4\pi i \, \p_{\tau} \log R=-E^{(1)}{}'(\xi )-
E^{(4)}{}'(\xi )+4\pi i \, \p_{\tau}
\log \Bigl ( \theta_2(0)
\theta_3(0)\Bigr ).
$$
The last term can be transformed using the heat equation 
(\ref{heat}) for theta-functions. Taking into account that 
$\theta_2'(0)=\theta_3'(0)=0$, we have:
$$
\begin{array}{lll}
4\pi i \, \p_{\tau} \log R & =& \displaystyle{-\, \p_x^2 
\log \Bigl (\theta_1(x|\tau )\, \theta_4(x|\tau )\Bigr )
\Bigr |_{x=\xi}\! +\, \p_x^2 \log 
\Bigl (\theta_2(x|\tau )\, \theta_3(x|\tau )\Bigr )
\Bigr |_{x=0}}
\\ && \\
& =& -\, \p_x^2 
\log \theta_1(x|\frac{\tau}{2})\Bigr |_{x=\xi} 
\! +\, \p_x^2 \log \theta_1(x |\frac{\tau}{2})\Bigr |_{x=\frac{1}{2}} 
\end{array}
$$
where the well known identities 
$$
\begin{array}{l}
2\theta_1(u|\tau )\theta_4(u|\tau )=
\theta_2(0|\frac{\tau}{2})\, \theta_1(u|\frac{\tau}{2}),
\\ \\
2\theta_2(u|\tau )\theta_3(u|\tau )=
\theta_2(0|\frac{\tau}{2})\, \theta_2(u|\frac{\tau}{2})
\end{array}
$$
are used. We see that the equality that we are going to prove,
i.e., 
$$
4\pi i \, \p_{\tau} \log R =(S'(\xi ))^2 =
\pi^2 \theta_4^4(0|\tau )\, 
\frac{\theta_2^2(\xi |\frac{\tau}{2})}{\theta_1^2(\xi |\frac{\tau}{2})}
$$
is equivalent to the identity
$$
\begin{array}{c}
-\, \p_x^2 
\log \theta_1(x|\frac{\tau}{2}) +\, \p_x^2 
\log \theta_1(x |\frac{\tau}{2})\Bigr |_{x=\frac{1}{2}} \, =\, 
\pi^2 \theta_4^4(0|\tau )\, \displaystyle{
\frac{\theta_2^2(x |\frac{\tau}{2})}{\theta_1^2(x |\frac{\tau}{2})}\,.}
\end{array}
$$
The latter is proved by the standard argument. The both sides 
are elliptic functions with periods $1$ and $\frac{\tau}{2}$ and
a pole of order 2 at $x=0$ of the form $x^{-2} +O(1)$
(to see this, one should use the identities (\ref{theta1prime}) and 
$\theta_4^2 (0|\tau )=\theta_3(0|\frac{\tau}{2})\,
\theta_4(0|\frac{\tau}{2})$).
Therefore,
their difference is a constant. Evaluating both sides at $x=\frac{1}{2}$,
we find that the constant is $0$.

\section*{Acknowledgements}

\addcontentsline{toc}{section}{Acknowledgements}

We thank I.Krichever, A.Levin, S.Natanzon, 
A.Orlov and T.Takebe for discussions.
Thanks are also due
to S.Kharchev whose notes \cite{Kharchev} on identities 
for theta-functions were very useful for us. 
Both authors were supported in part by RFBR grant
14-02-00627. The work of A.Z. was also supported in part
by grant NSh-1500.2014.2 
for support of leading scientific schools.

\end{document}